\documentclass[10pt,conference]{IEEEtran}

\usepackage[noadjust]{cite}
\usepackage{ifthen}
\newboolean{blind}   
\setboolean{blind}{false}   

\usepackage{xcolor}
\usepackage{colortbl}
\usepackage{booktabs}
\usepackage{graphicx}
\usepackage{multirow}
\usepackage{pgfplots}
\usepackage{ifthen}
\usepackage{subcaption}
\usepackage{amsmath,amsthm,amssymb,amsxtra} 
\usepackage{mathtools}
\usepackage{float} 
\usepackage{hyperref}
\usepackage{pdflscape}

\usepgfplotslibrary{colormaps}
\pgfplotsset{compat=newest} 
\pgfdeclarelayer{background}
\pgfdeclarelayer{background_1}
\pgfdeclarelayer{background_2}
\pgfdeclarelayer{background_3}
\pgfdeclarelayer{foreground_6}
\pgfdeclarelayer{foreground_5}
\pgfdeclarelayer{foreground_4}
\pgfdeclarelayer{foreground_3}
\pgfdeclarelayer{foreground_2}
\pgfdeclarelayer{foreground_1}
\pgfdeclarelayer{foreground}
\pgfsetlayers{background,background_1,background_2,background_3,main,foreground_6,foreground_5,foreground_4,foreground_3,foreground_2,foreground_1,foreground}

\usepackage{tikz}
\usetikzlibrary{arrows, arrows.meta, positioning, quotes, shapes, fit, calc, decorations.pathreplacing, calligraphy, plotmarks, external}
\tikzstyle{block} = [draw, thick, text=black, align=center, minimum width=1.2cm, minimum height=0.6cm, font=\small]
\tikzstyle{Block} = [draw, thick, rounded corners, text=black, align=center, minimum width=2cm, minimum height=1.2cm, font=\small]
\usepackage{circuitikz}
\ctikzset{bipoles/cuteswitch/thickness=0.3}

\usepackage{amsmath}
\usepackage[binary-units=true]{siunitx}
\DeclareSIUnit\BL{\giga\bit\kilo\meter\per\second}
\DeclareSIUnit\FL{\decibel\per\kilo\meter}
\DeclareSIUnit\CD{\pico\second\per\nano\meter\per\kilo\meter}

\usepackage{mleftright}
\usepackage{relsize}
\usepackage{svg}
\usepackage{textgreek}
\usepackage{bm}
\usepackage{verbatim}
\usepackage{tabularray}

\usepackage[nohyperlinks,nolist]{acronym}

\acrodef{conv1d}[conv1d]{one-dimensional convolution}

\acrodef{ai}[AI]{artificial intelligence}
\acrodef{ask}[ASK]{amplitude-shift keying}
\acrodef{awgn}[AWGN]{additive white Gaussian noise}
\acrodef{ann}[ANN]{artificial neural network}
\acrodef{asic}[ASIC]{application-specific integrated circuit}

\acrodef{bce}[BCE]{binary cross-entropy}
\acrodef{ber}[BER]{bit error ratio}
\acrodef{bler}[BLER]{block error ratio}
\acrodef{bpsk}[BPSK]{binary phase-shift keying}
\acrodef{bram}[BRAM]{block random access memory}
\acrodef{bilstm}[biLSTM]{bidirectional long short-term memory}
\acrodef{bp}[BP]{backward pass}

\acrodef{cd}[CD]{chromatic dispersion}
\acrodef{cir}[CIR]{combined impulse response}
\acrodef{cma}[CMA]{constant modulus algorithm}
\acrodef{cnn}[CNN]{convolutional neural network}
\acrodef{cpu}[CPU]{central processing unit}
\acrodef{ctp}[CTP]{channel transition probability}
\acrodefplural{ctp}[CTPs]{channel transition probabilities}
\acrodef{csi}[CSI]{channel state information}

\acrodef{dnn}[DNN]{deep neural network}
\acrodef{dsp}[DSP]{digital signal processor}
\acrodef{dram}[DRAM]{dynamic random access memory}
\acrodef{dfe}[DFE]{decision-feedback equalizer}
\acrodef{dop}[DOP]{degree of paralellism}
\acrodef{dbp}[DBP]{digital back-propagation}

\acrodef{elu}[ELU]{exponential linear unit}
\acrodef{ecnn}[ECNN]{enhanced convolutional neural network}

\acrodef{fc}[FC]{fully connected}
\acrodef{fec}[FEC]{forward error correction}
\acrodef{fir}[FIR]{finite impulse response}
\acrodef{fifo}[FIFO]{first in first out}
\acrodef{fpga}[FPGA]{field programmable gate array}
\acrodef{fp}[FP]{forward pass}
\acrodef{ff}[FF]{flip-flop}

\acrodef{gan}[GAN]{generative adversarial network}
\acrodef{gpu}[GPU]{graphics processing unit}

\acrodef{hls}[HLS]{high-level synthesis}
\acrodef{hwa}[HWA]{historical weight averaging}

\acrodef{imdd}[IM/DD]{intensity modulation with direct detection}
\acrodef{iot}[IoT]{Internet of things}
\acrodef{isi}[ISI]{inter-symbol interference}

\acrodef{ldpc}[LDPC]{low-density parity-check}
\acrodef{lms}[LMS]{least-mean-square}
\acrodef{lut}[LUT]{look-up table}

\acrodef{mlsd}[MLSD]{maximum-likelihood sequence detection}
\acrodef{mse}[MSE]{mean-squared-error}
\acrodef{mlse}[MLSE]{maximum likelihood sequence estimator}
\acrodef{mac}[MAC]{multiply-accumulate}
\acrodef{ma}[MA]{moving average}
\acrodef{map}[MAP]{maximum a posteriori}
\acrodef{msm}[MSM]{merge stream module}
\acrodef{mzm}[MZM]{mach-zehnder modulator}
\acrodef{nn}[NN]{neural network}
\acrodef{ns}[NS]{non-saturating}

\acrodef{ogm}[OGM]{overlap generate module}
\acrodef{orm}[ORM]{overlap remove module}
\acrodef{onu}[ONU]{optical network unit}
\acrodef{olt}[OLT]{optical line terminal}
\acrodef{odn}[ODN]{optical distribution network}

\acrodef{pam}[PAM]{pulse-amplitude modulation}
\acrodef{pdf}[pdf]{probability density function}
\acrodef{pe}[PE]{processing element}
\acrodef{pmf}[pmf]{probability mass function}
\acrodef{pon}[PON]{passive optical network}

\acrodef{qlf}[QLF]{quantization loss factor}
\acrodef{qam}[QAM]{quadrature amplitude modulation}

\acrodef{rc}[RC]{raised-cosine}
\acrodef{relu}[ReLU]{rectified linear unit}
\acrodef{rls}[RLS]{recursive least squares}
\acrodef{rnn}[RNN]{recurrent neural network}

\acrodef{ser}[SER]{symbol error rate}
\acrodef{sgd}[SGD]{stochastic gradient descent}
\acrodef{sld}[SLD]{square-law detection}
\acrodef{snr}[SNR]{signal-to-noise ratio}
\acrodef{sps}[sps]{samples per symbol}
\acrodef{ssmf}[SSMF]{standard single-mode fiber}
\acrodef{simd}[SIMD]{single instruction multiple data}
\acrodef{ssm}[SSM]{split stream module}
\acrodef{spb}[SPB]{symbols per batch}
\acrodef{slda}[SLDA]{streaming linear discriminant analysis}
\acrodef{svm}[SVM]{support vector machine}

\acrodef{ti}[TI]{training iteration}
\acrodef{tpdr}[TPDR]{true-positive detection rate}
\acrodef{tpu}[TPU]{tensor processing unit}

\acrodef{vnle}[VNLE]{Volterra-based nonlinear equalizer}

\acrodef{fpr}[FPR]{false positive rate}

\definecolor{RPTU_BlueGray}{RGB}{80,114,137}
\definecolor{RPTU_GreenGray}{RGB}{119,182,186}
\definecolor{RPTU_DarkBlue}{RGB}{4,44,88}
\definecolor{RPTU_LightBlue}{RGB}{106,178,231}
\definecolor{RPTU_DarkGreen}{RGB}{0,107,107}
\definecolor{RPTU_LightGreen}{RGB}{38,208,124}
\definecolor{RPTU_Violett}{RGB}{76,53,117}
\definecolor{RPTU_Pink}{RGB}{209,56,150}
\definecolor{RPTU_Red}{RGB}{227,27,76}
\definecolor{RPTU_Orange}{RGB}{255,162,82}
\definecolor{RPTU_Black}{RGB}{0,0,0}
\definecolor{RPTU_White}{RGB}{255,255,255}

\AtBeginDocument{%
  \providecommand\BibTeX{{%
    \normalfont B\kern-0.5em{\scshape i\kern-0.25em b}\kern-0.8em\TeX}}}

\begin{document}

\title{ECNN: A Low-complex, Adjustable CNN for Industrial Pump Monitoring Using Vibration Data}

\ifthenelse{\boolean{blind}}
{
\author{\normalfont \textit{Authors omitted due to double-blind review}}
}
{

\author{
\IEEEauthorblockN{Jonas Ney and Norbert Wehn}
\IEEEauthorblockA{\textit{Microelectronic Systems Design (EMS)}, \textit{RPTU Kaiserslautern-Landau}, Germany \\
\{\texttt{jonas.ney}, \texttt{norbert.wehn}\}\texttt{@rptu.de}} 
}
}

\maketitle


\begin{abstract}

Industrial pumps are essential components in various sectors such as manufacturing, energy production, and water treatment, where their failures can cause significant financial and safety risks. Anomaly detection can be used to reduce those risks and increase reliability. In this work, we propose a novel \ac{ecnn} to predict the failure of an industrial pump based on the vibration data captured by an acceleration sensor. The \ac{cnn} is designed with a focus on low complexity to enable its implementation on edge devices with limited computational resources. Therefore, a detailed design space exploration is performed to find a topology satisfying the trade-off between complexity and accuracy. Moreover, to allow for adaptation to unknown pumps, our algorithm features a pump-specific parameter that can be determined by a small set of normal data samples. Finally, we combine the \ac{ecnn} with a threshold approach to further increase the performance and satisfy the application requirements. As a result, our combined approach significantly outperforms a traditional statistical approach and a classical \ac{cnn} in terms of accuracy. 
To summarize, this work provides a novel, low-complex, \ac{cnn}-based algorithm that is enhanced by classical methods to offer high accuracy for anomaly detection of industrial pumps. 
\end{abstract}

\ifthenelse{\boolean{blind}}
{
\let\thefootnote\relax\footnotetext{\centering \large \textit{Grant agreements omitted due to double-blind review}}
}
{
\let\thefootnote\relax\footnotetext{This work was funded by the German Federal Ministry of Education and Research (BMBF) under grant agreement 16KIS1662 (SIPSENSIN).  Further, it was funded by the Carl Zeiss Stiftung under the Sustainable Embedded AI project (P2021-02-009). We sincerely thank our partner KSB for providing the dataset and the application requirements.}
}

\acresetall

\section{Introduction}
\label{sec:introduction}

Industrial pumps are essential components in sectors ranging from manufacturing and energy production to water treatment and chemical processing. The failure of a pump can lead to significant downtime, increased maintenance costs, and even catastrophic system breakdowns, posing substantial financial and safety risks. According to the Hydraulic Institute, around \SI{25}{\percent} of the total lifespan cost of an industrial pump can be attributed to maintenance and repair~\cite{pumpsorglifecycle}. Therefore, ensuring the reliable operation of pumps is crucial for both economic and safety reasons. 

Anomaly detection describes the process of identifying unusual patterns or outliers in data that do not conform to expected behavior. For industrial pumps, anomaly detection can be applied to the vibration data of the pump captured by an acceleration sensor. This way, potential faults and malfunctions can be identified at an early stage, enabling timely intervention and preventing unexpected failures.

Vibration analysis is a well-established method for monitoring the health of rotating machinery, including pumps~\cite{rao2010MechanicalVibrations, adams2009RotatingMachineryVibration}. Changes in vibration patterns can indicate various issues such as imbalance, misalignment, and other mechanical problems. Traditionally, the algorithms used for anomaly detection can be categorized into statistical methods~\cite[Chapter~4.1.1]{JARDINE20061483} and classical machine learning methods like k-nearest-neighbors~\cite{tian2014anomaly, lei2020anomaly} or \ac{svm}~\cite{amer2013}. Additionally, with the rise of deep learning, \ac{nn}-based methods became more and more popular for anomaly detection in recent years~\cite{zhang2020} driven by their state-of-the-art performance in time-series tasks such as natural language processing~\cite{hinton2012} or speech recognition~\cite{bert}. However, the novel \ac{nn}-based methods commonly introduce a high computational complexity and large memory footprint, whereas anomaly detection algorithms are often implemented on battery-powered edge devices with limited computational resources. Thus, in this work, we present a novel algorithm for \ac{nn}-based anomaly detection of pump vibration data using a low-complexity \ac{nn}, developed through extensive design space exploration. 

Since our dataset consists of multiple pumps with varying characteristics, one goal of this work is to provide an algorithm that is applicable to a diverse set of different pumps without major modifications. This challenge is further complicated by the fact that only \textit{normal} samples are available for fine-tuning as newly manufactured pumps are assumed to operate normally. 
Therefore, we design a \ac{nn} architecture, referred to as \ac{ecnn}, which includes an adaptable parameter that can be adjusted to each pump individually without relying on retraining of the \ac{nn}. Further, we present an algorithm to estimate this pump-specific parameter based on \textit{normal} samples only. Moreover, we show that this approach provides much better performance than a conventional non-adjustable \ac{cnn}. Finally, we show how the \ac{ecnn} can be combined with a conventional threshold approach to achieve even higher accuracy. 

\let\thefootnote\relax\footnotetext{
The source code of the models as well as the training and testing flow is available at: \url{https://github.com/jney-eit/ECNN_SSCI}}

\section{System Overview}
\label{sec:system_overview}

The main goal of this work is to provide an accurate algorithm for the detection of anomalies in the vibration data of industrial pumps. In the following, we describe the dataset on which our evaluations are based and discuss the requirements of the application.

\subsection{Dataset}
\ifthenelse{\boolean{blind}}
{
Our experiments are based on a custom pump anomaly detection dataset of the company \textit{COMPANY NAME OMITTED DUE TO DOUBLE-BLIND REVIEW.} 
}
{
Our experiments are based on a custom pump anomaly detection dataset of the international pump manufacturing company \textit{KSB}~\cite{KSBWebsite}.
}
 The dataset consists of three-dimensional vibration data of acceleration sensors placed on different industrial pumps. Each sample contains a \textit{pump id} defining the specific pump, the vibration data of size $800 \times 3$, and a label \textit{normal} or \textit{abnormal}. Overall the dataset contains \num{633} different pumps with a total of \num{377676} samples.

For our experiments, only the pumps are considered that contain normal as well as abnormal samples. This results in a dataset of \num{108} pumps with a total of \num{251025} samples. Overall this dataset consists of \num{83699} normal and \num{167326} abnormal samples. 
Mathematically, the dataset is a set of $n$ pumps
\begin{align*}
    P = \{p_1, p_2, \dots, p_n\}
\end{align*}
and contains $m$ samples for each pump $p_i$: 
\begin{align*}
    S_i = \{s_{i1}, s_{i2}, \dots, s_{im}\} \;   .
\end{align*}
Each sample $s_{ij}$ consists of input $X_{ij}$ and label $Y_{ij}$ with
\begin{align*}
    X_{ij} = \{x_{ij}, y_{ij}, z_{ij}\}
\end{align*}
where $x_{ij}$, $y_{ij}$, and $z_{ij}$ are vectors of \num{800} elements respectively, and 
\vspace{-4mm}
\begin{align*}
    Y_{ij} \in \{0, 1\} \;   .
\end{align*}


\subsection{Application Requirements}

The application requires the algorithm to be applied to newly manufactured pumps where no labeled samples are available beforehand. This fact imposes constraints on how the dataset should be partitioned for training and testing. The algorithm will eventually applied to samples of unseen pumps. Thus, the training dataset should not contain any samples of pumps that are used for testing. This ensures that an algorithm with high testing accuracy is able to generalize well to unseen pumps. 
However, we are allowed to perform fine-tuning for the pumps of the test dataset. It is assumed that a newly manufactured pump functions normally at first use. Therefore, normal samples of this pump are available for fine-tuning the previously trained algorithm. Consequently, characteristics of previously unseen pumps can be learned in an adaptation phase from normal data samples.

\ifthenelse{\boolean{blind}}
{
According to our partner \textit{COMPANY NAME OMITTED DUE TO DOUBLE BLIND REVIEW}, for an industrial use-case, an accuracy of \SI{85}{\percent} should be achieved in combination with a \ac{tpdr} of \SI{98}{\percent} where the \ac{tpdr} is given as the percentage of pumps for which at least one sample is correctly identified as positive.
}
{
According to our partner \textit{KSB}, for an industrial use-case, an accuracy of \SI{85}{\percent} should be achieved in combination with a \ac{tpdr} of \SI{98}{\percent} where the \ac{tpdr} is given as the percentage of pumps for which at least one sample is correctly identified as positive.
}

Since the algorithm will be applied a posteriori to pumps that are already in use where modifications to the power circuit are impractical, it should be suited for deployment on battery-powered edge devices. This imposes constraints on the complexity of the algorithm as higher computational complexity results in increased power consumption and reduced energy efficiency. Therefore, during design space exploration, it is crucial to consider the trade-off between computational complexity and accuracy.  
\section{Algorithms}
\label{sec:algorithms}

In the following, we describe the algorithms that are evaluated for detecting anomalies in the vibration data of the industrial pumps. All approaches receive the three-dimensional vibration data as input and predict if the processed sample contains anomalies or corresponds to a pump functioning normally. 

\subsection{Threshold Approach}

The first algorithm we evaluate is a classical threshold approach. For this approach, as a first step, the mean of the normal samples of each pump is determined in $x$, $y$, and $z$ dimensions:
\begin{align*}
    \mu_i^0 &= \{\mu_{xi}^0, \mu_{yi}^0, \mu_{zi}^0\}\\
    \mu_{di}^0 &= \underset{j,k}{\mathrm{mean}}(d_{ijk} | Y_{ij} = 0), \quad d \in {x, y, z} \;   .
\end{align*}
To perform a prediction for an unseen test sample, the \ac{mse} between the mean $\mu_i^0$ and every datapoint of the new sample is calculated:

\begin{equation}
    \epsilon_{ic} = \frac{1}{3} \biggl\Vert (X_{ij} - \mu_i^0)^2 \biggr\Vert_1 \;   .
\end{equation}

If $\epsilon_{ic} < T_i$ for a pump-specific threshold $T_i$ we classify the sequence as \textit{normal}, otherwise as \textit{abnormal}. The pump-specific threshold needs to be determined based on the \textit{normal} samples of the unseen pump. The key idea of this approach is that faulty pumps have vibrations with a higher amplitude and therefore a higher \ac{mse} with respect to the normal samples. 

\subsection{Neural-Network-Based Approaches}
\label{subsec:cnn}

As a second category of algorithms, we evaluate different \ac{nn}-based approaches. In general, \acp{nn} are well suited for finding hidden structures in large sets of data. Particularly \acp{cnn} are commonly used for processing one-dimensional sequential data, as their convolutional layers can efficiently capture local patterns and temporal dependencies.

\subsubsection{CNN Template}

As an adjustable template for our \ac{cnn}, we select the following topology: the \ac{cnn} is composed of $L$ convolutional layers with identical kernel size $K$. Each convolutional layer but the last is followed by batch normalization and \ac{relu} activation functions. Three channels are used for the input sequence, while subsequent activations consist of $C$ channels. The last convolutional layer outputs one channel and is followed by a global average pooling layer to produce a single output value. The sample is predicted to be \textit{normal} if the output value is $< 0.5$ and \textit{abnormal} otherwise. In the following, this simple \ac{cnn} architecture is referred to as \textit{default CNN}.

We use a parametrizable \ac{cnn} template to be able to explore the complexity-accuracy trade-off by training different \ac{cnn} configurations. This way we can find a low-complex model that still achieves sufficient accuracy. 

\subsubsection{Enhanced CNN}

Besides the default \ac{cnn}, we also evaluate an \ac{ecnn} that receives an additional input inspired by the threshold approach. The additional input is given as $A_{ic} = \{a_{xic}, a_{yic}, a_{zic} \}$ for the pump $c$. In particular, for a sample $X_{ic}$, the normal mean $\mu_i^0$ is subtracted from each datapoint, and the result is multiplied by a pump-specific factor $F_i$:
\begin{equation}
    \label{eq:Aic}  
    A_{ic}
    =
    \begin{pmatrix}
        a_{xic} \\
        a_{yic} \\
        a_{zic}
    \end{pmatrix}
    =
    F_i \cdot 
    \begin{pmatrix}
        x_{ick} - \mu_{xi}^0 \\
        y_{ick} - \mu_{yi}^0 \\
        z_{ick} - \mu_{zi}^0
    \end{pmatrix}
\end{equation}

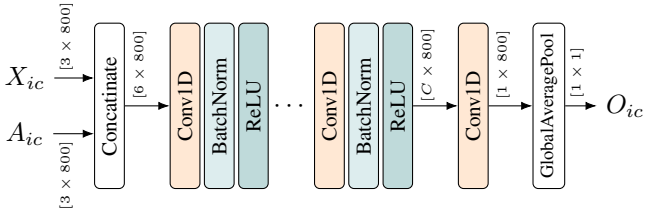
\begin{figure}[h]
    \centering
    \vspace{-0.3cm}
	\tikzsetnextfilename{CNN_final_topology}
\begin{tikzpicture}[node distance=0.2,>=latex]
\def\minHei{1mm} \def\minWid{2.2cm}

\node (X) {$X_{ic}$};
\node[below=0.2cm of X] (A) {$A_{ic}$}; 

\path let \p1 = ($(X)!0.5!(A)$) in node (Middle) at (\x1, \y1) {};

\node[inner sep=1mm, block, thin, right=1cm of Middle.east, anchor=center, rotate=90, font=\footnotesize, minimum height=\minHei, minimum width=\minWid, rounded corners=2pt] (Concat) {Concatinate};

\node[inner sep=1mm, block, thin, right=0.8cm of Concat.south, anchor=center, rotate=90, font=\footnotesize, minimum height=\minHei, minimum width=\minWid, fill=RPTU_Orange!25, rounded corners=2pt] (Conv1D_1) {Conv1D};

\node[inner sep=1mm, block, thin, right=0.25cm of Conv1D_1.south, anchor=center, rotate=90, font=\footnotesize, minimum height=\minHei, minimum width=\minWid, fill=RPTU_GreenGray!25, rounded corners=2pt] (BatchNorm_1) {BatchNorm};

\node[inner sep=1mm, block, thin, right=0.25cm of BatchNorm_1.south, anchor=center, rotate=90, font=\footnotesize, minimum height=\minHei, minimum width=\minWid, fill=RPTU_DarkGreen!25, rounded corners=2pt] (Relu_1) {ReLU};

\node[right=0.3cm of Relu_1.south, anchor=center, font=\normalsize, minimum height=\minHei, minimum width=\minWid] (Dots) {$\cdots$};

\node[inner sep=1mm, block, thin, right=-0.6cm of Dots.east, anchor=center, rotate=90, font=\footnotesize, minimum height=\minHei, minimum width=\minWid, fill=RPTU_Orange!25, rounded corners=2pt] (Conv1D_2) {Conv1D};

\node[inner sep=1mm, block, thin, right=0.25cm of Conv1D_2.south, anchor=center, rotate=90, font=\footnotesize, minimum height=\minHei, minimum width=\minWid, fill=RPTU_GreenGray!25, rounded corners=2pt] (BatchNorm_2) {BatchNorm};

\node[inner sep=1mm, block, thin, right=0.25cm of BatchNorm_2.south, anchor=center, rotate=90, font=\footnotesize, minimum height=\minHei, minimum width=\minWid, fill=RPTU_DarkGreen!25, rounded corners=2pt] (Relu_2) {ReLU};

\node[inner sep=1mm, block, thin, right=0.8cm of Relu_2.south, anchor=center, rotate=90, font=\footnotesize, minimum height=\minHei, minimum width=\minWid, fill=RPTU_Orange!25, rounded corners=2pt] (Conv1D_3) {Conv1D};

\node[inner sep=1mm, block, thin, right=0.8cm of Conv1D_3.south, anchor=center, rotate=90, font=\scriptsize, minimum height=\minHei, minimum width=\minWid, rounded corners=2pt] (GlobalAveragePool) {GlobalAveragePool};

\node[right=0.8cm of GlobalAveragePool.south, anchor=center] (Output) {$O_{ic}$};

\draw[-{Latex[length=1.5mm]}] (X) -- node[midway, above, font=\tiny, anchor=center, rotate=90, xshift=0.55cm, yshift=0.1cm] {$[3 \times 800]$} (X -| Concat.north);
\draw[-{Latex[length=1.5mm]}] (A) -- node[midway, below, font=\tiny, anchor=center, rotate=90, xshift=-0.55cm, yshift=0.1cm] {$[3 \times 800]$} (A -| Concat.north);

\draw[-{Latex[length=1.5mm]}] (Concat.south) -- node[midway, above, font=\tiny, anchor=center, rotate=90, xshift=0.55cm, yshift=0.1cm] {$[6 \times 800]$} (Conv1D_1.north);

\draw[-{Latex[length=1.5mm]}] (Relu_2.south) -- node[midway, above, font=\tiny, anchor=center, rotate=90, xshift=0.55cm, yshift=0.1cm] {$[C \times 800]$} (Conv1D_3.north);

\draw[-{Latex[length=1.5mm]}] (Conv1D_3.south) -- node[midway, above, font=\tiny, anchor=center, rotate=90, xshift=0.55cm, yshift=0.1cm] {$[1 \times 800]$} (GlobalAveragePool.north);

\draw[-{Latex[length=1.5mm]}] (GlobalAveragePool.south) -- node[midway, above, font=\tiny, anchor=center, rotate=90, xshift=0.45cm, yshift=0.05cm] {$[1 \times 1]$}(Output.west);

\end{tikzpicture}
	\caption{Topology of the \ac{ecnn}}
	\label{fig:ecnn_topology}
\end{figure}

Afterwards $A_{ic}$ is concatenated with the current input $X_{ic}$ to form a feature map of length $800$ with $6$ channels. This input is passed to the \ac{cnn} to perform a prediction. The topology is shown in Fig. \ref{fig:ecnn_topology}. The pump-specific factor $F_i$ needs to be determined based on the \textit{normal} samples of the unseen pump. 

\vspace{-0.3cm}
\subsection{Combined Approach}
For our experiments, we also evaluate an approach that combines the thresholding algorithm with the \ac{ecnn}. For this approach, the \ac{fpr} of the \ac{ecnn} is evaluated for a specific pump. Based on the result, either the \ac{ecnn} or the threshold algorithm is used for further predictions for this pump. This approach is described in more detail in the results section. 
\section{Results}
\label{sec:results}

For the \ac{nn} results presented in this section, the training was performed using PyTorch on a server with an Nvidia V100 \ac{gpu}. We trained the models for \num{100} epochs with a learning rate of \num{0.001}, Adam optimizer, and \ac{mse} loss. 

\subsection{Neural Network Design Space Exploration}

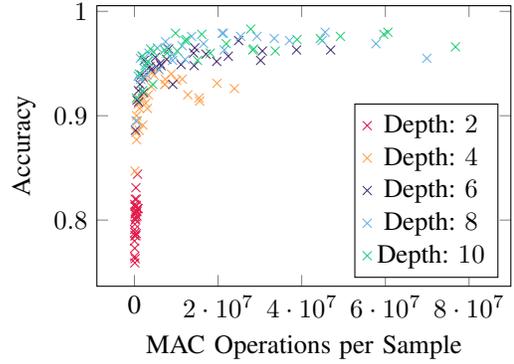
\begin{figure}[t]
	\centering
	\tikzsetnextfilename{CNN_DSE}

\begin{tikzpicture}
    []

	\begin{axis}[
            xmax=90000000,
            legend pos=south east,
            xlabel={MAC Operations per Sample},
            ylabel={Accuracy},
            width=0.8*\columnwidth,
            height=0.6*\columnwidth,
            scaled x ticks=false,
            xticklabels={0, \num{2e7}, \num{4e7}, \num{6e7}, \num{8e7}},
            xtick={0, 20000000,40000000,60000000,80000000},
            ]

    \addplot [only marks, mark=x, color=RPTU_Red, mark options={scale=1.2}] table [x expr=\thisrow{MAC}*800, y=Accuracy] {data/nn_dse_depth2.txt};
    \addlegendentry{Depth: \num{2}};

    \addplot [only marks, mark=x, color=RPTU_Orange, mark options={scale=1.2}] table [x expr=\thisrow{MAC}*800, y=Accuracy] {data/nn_dse_depth4.txt};
    \addlegendentry{Depth: \num{4}};

    \addplot [only marks, mark=x, color=RPTU_Violett, mark options={scale=1.2}] table [x expr=\thisrow{MAC}*800, y=Accuracy] {data/nn_dse_depth6.txt};
    \addlegendentry{Depth: \num{6}};

    \addplot [only marks, mark=x, color=RPTU_LightBlue, mark options={scale=1.2}] table [x expr=\thisrow{MAC}*800, y=Accuracy] {data/nn_dse_depth8.txt};
    \addlegendentry{Depth: \num{8}};

    \addplot [only marks, mark=x, color=RPTU_LightGreen, mark options={scale=1.2}] table [x expr=\thisrow{MAC}*800, y=Accuracy] {data/nn_dse_depth10.txt};
    \addlegendentry{Depth: \num{10}};
    
    \end{axis}

\end{tikzpicture}
	\caption{Design space exploration of different \ac{nn} architectures. The x-axis gives the complexity in terms of \ac{mac} operations of the \ac{nn} to process one sample. The y-axis gives the test accuracy of the \ac{nn}.}
	\label{fig:dse}
    \vspace{-2mm}
\end{figure}

In general, the design of \ac{nn}-based algorithms is associated with a huge set of hyperparameters, spanning an enormously large design space nearly infeasible to explore completely. Therefore, we constrain our design space to the topology template presented in Sec. \ref{subsec:cnn}. This template is utilized to find a topology with a good trade-off between computational complexity and accuracy. 
For this exploration, we focus on the default \ac{cnn} where the best configurations are used for the \ac{ecnn} and the combined approach later on. In particular, our search space is spanned by the depth of the network $D$, the kernel size $K$, and the number of channels $C$.
For exploring the \ac{nn} topology, we divide the dataset into a fixed training and test set instead of performing cross-validation for each pump. We avoid individual training and testing for each pump configuration as it would immensely increase the exploration time, making it impractical. This has the drawback that samples of each pump are contained in the training set, which causes a difference in accuracy as compared to the pump-specific cross-validation. However, our experiments indicate that this approach offers a reliable estimate of the final accuracy, sufficient for comparing different configurations.
The result of this design space exploration is shown in Fig. \ref{fig:dse}.

Each point of the plot corresponds to a different \ac{cnn} configuration. It can be seen that for most models, a higher complexity leads to a higher accuracy. In particular, there is a steep increase in accuracy from \num{0} to \num{20}k \ac{mac} operations. For higher complexity, the accuracy increases only slightly. For further exploration, the most promising Pareto optimal models were selected from the design space exploration results. 

\subsection{Cross-Validation}
\label{subsec:cross_validation}

For a more detailed evaluation of the different algorithms and \ac{nn} configurations, we perform $n$-fold cross-validation where $n=108$, which is the number of different pumps. In particular, the training dataset for pump $i$ consists of all samples of the remaining $107$ pumps, and the test dataset is given by all samples of pump $i$. Each \ac{nn} configuration is individually trained and tested for each pump. The final accuracy is given as the average accuracy of all pumps. This way, the generalization capability of the \ac{nn} with respect to unseen pumps can be evaluated. 

We performed this cross-validation for the \ac{cnn}, the \ac{ecnn} and the threshold algorithm. For the \ac{ecnn} and the threshold algorithm, the pump-specific parameters $T_i$ and $F_i$ are selected optimally based on labeled data. This optimal selection is not applicable in practice since the labels for unseen pumps are not known. In Sec. \ref{subsec:pump_specific_parameter_selection}, we present an approach on how the optimal pump-specific parameter can be estimated in practice and show how a non-optimal parameter affects the accuracy. 
Tab. \ref{tab:cross_validation} shows the cross-validation results for different models of the \ac{cnn}, the \ac{ecnn}, and the threshold algorithm. The accuracy corresponds to the $n$-fold cross-validation accuracy and the \ac{tpdr} gives the percentage of pumps where at least one true positive sample is detected in the test set. The cells that satisfy our application requirements of \SI{85}{\percent} accuracy and \SI{98}{\percent} \ac{tpdr} are highlighted in green. In the table, it is shown that our \acp{ecnn} outperforms all default \acp{cnn} of similar size in terms of accuracy and \ac{tpdr}. It can be seen that in general, the accuracy increases with higher complexity. However, the accuracy starts to saturate at around \SI{90}{\percent}, beyond this point, it can't be increased significantly even with much more complex models. Further, it can be seen that the threshold algorithm provides competitive performance. In the table, we also highlight the least complex \ac{ecnn} that satisfies the application requirements, which is selected for further evaluation in the following.  
\newcommand\tikzmark[1]{%
  \tikz[remember picture,overlay] \node (#1) {};}
\begin{table}[t]
\centering
\caption{Cross-Validation Results for different \ac{nn} architectures}
\label{tab:cross_validation}
\begin{tabular}{ccccccc}
\toprule
\multirow{2}{*}{Alg.} & \multirow{2}{*}{Depth} & Kernel & \multirow{2}{*}{Channels} & MAC & Acc. & \acs{tpdr} \\
&  & Size & & Operations & (\si{\percent}) & (\si{\percent}) \\
\midrule
CNN & \num{4} & \num{11} & \num{5} & \num{6.2e5} & \num{71.3} & \num{89.8} \\
CNN & \num{6} & \num{23} & \num{5} & \num{2.2e6} & \num{71.3} & \num{90.7} \\
CNN & \num{10} & \num{23} & \num{10} & \num{1.5e7} & \num{72.7} & \num{91.7} \\
CNN & \num{10} & \num{19} & \num{20} & \num{5.0e7} & \num{73.4} & \num{91.7} \\
CNN & \num{30} & \num{23} & \num{30} & \num{4.7e8} & \num{72.8} & \num{90.7} \\
\midrule
ECNN & \num{4} & \num{11} & \num{5} & \num{7.5e5} & \cellcolor{RPTU_LightGreen!25} \num{86.8} & \num{97.2} \\
ECNN & \num{6} & \num{23} & \num{5} & \num{2.5e6} & \cellcolor{RPTU_LightGreen!25} \num{86.9} & \num{97.2} \\
\tikzmark{a} ECNN & \num{10} & \num{23} & \num{10} & \num{1.6e7} & \cellcolor{RPTU_LightGreen!25} \num{88.0} & \cellcolor{RPTU_LightGreen!25} \num{98.1} \tikzmark{b}\\
ECNN & \num{10} & \num{19} & \num{20} & \num{5.1e7} & \cellcolor{RPTU_LightGreen!25} \num{89.4} & \cellcolor{RPTU_LightGreen!25} \num{98.1} \\
ECNN & \num{30} & \num{23} & \num{30} & \num{4.7e8} & \cellcolor{RPTU_LightGreen!25} \num{90.3} & \num{96.3} \\
\midrule
\multicolumn{5}{c}{Threshold} & \cellcolor{RPTU_LightGreen!25} \num{88.7} & \num{96.3} \\
\bottomrule
\end{tabular}
\end{table}
\begin{tikzpicture}[remember picture,overlay]
\draw[line width=1pt,draw=RPTU_LightBlue,rounded corners=1pt]
 ([xshift=-4pt,yshift=3.6pt]a.north) rectangle ([xshift=5.5pt,yshift=1pt]b.south);
\end{tikzpicture}

\subsection{Pump-Specific Parameter Selection}
\label{subsec:pump_specific_parameter_selection}

The results of Sec. \ref{subsec:cross_validation} are based on optimal pump-specific parameters $T_i$ and $F_i$, selected based on labeled data. In practice, this approach is not applicable since only normal samples can be used for fine-tuning. Thus, we provide a method to select the parameters based on the \ac{fpr}. Therefore, as an initialization step for a pump $i$, only normal samples are fed to the model and the parameter $F_i$ is set to a high value. This results in a high \ac{fpr}. Afterwards, the pump-specific parameter is slowly reduced while tracking the \ac{fpr}. If the \ac{fpr} becomes smaller than \SI{10}{\percent}, the pump-specific parameter is fixed to the current value. A similar method is applied for $T_i$. This way, for most pumps, the parameters are close to their 
 optimal value since a low \ac{fpr} often corresponds to a high accuracy.  

 As a fourth method besides the \ac{cnn}, the \ac{ecnn}, and the threshold algorithm, we give the results for a combined approach. This approach combines the \ac{ecnn} and the threshold algorithm based on the \ac{fpr} of the \ac{ecnn} in the following way: first, it is searched for the pump-specific parameter $F_i$ of the \ac{ecnn}. If an \ac{fpr} $<\SI{10}{\percent}$ is achieved, the corresponding parameter is used with the \ac{ecnn} model for this pump. However, for some pumps, the \ac{fpr} always stays above \SI{10}{\percent} with the \ac{ecnn}. In this case, the threshold algorithm is used for this pump. 
 
 The results are shown in Tab. \ref{tab:algorithm_comparison}, where \textit{Optimal} corresponds to the optimal specific-pump parameter, for \textit{Fixed} the parameter is not adjusted at all, and \textit{FPR} uses the previously described approach. Again, we highlight the cells that satisfy our application requirements.

\begin{table}[t]
\centering
\caption{Pump-specific Parameter Selection Results}
\label{tab:algorithm_comparison}
\begin{tabular}{cccc}
\toprule
Alg. & Param. Selection & Acc. (\si{\percent}) & \acs{tpdr}  (\si{\percent}) \\
\midrule
CNN & -- & \num{73.4} & \num{91.7} \\
\midrule
ECNN & Optimal & \cellcolor{RPTU_LightGreen!25} \num{89.4} & \cellcolor{RPTU_LightGreen!25} \num{98.1} \\
ECNN & Fixed & \num{73.0} & \num{97.2} \\
ECNN & FPR & \num{81.5} & \cellcolor{RPTU_LightGreen!25} \num{99.1} \\
\midrule
Threshold & Optimal & \cellcolor{RPTU_LightGreen!25} \num{88.7} & \num{96.3} \\
Threshold & Fixed & \num{66.5} & \num{97.2} \\
Threshold & FPR & \num{82.6} & \cellcolor{RPTU_LightGreen!25} \num{99.1} \\
\midrule
Combined & FPR & \cellcolor{RPTU_LightGreen!25} \num{86.9} & \cellcolor{RPTU_LightGreen!25} \num{99.1} \\
\bottomrule
\end{tabular}
\end{table}

It can be seen, that the selection of the pump-specific parameter highly influences the accuracy of the algorithm. For the \ac{cnn} the accuracy is reduced to only \SI{73}{\percent} and for the threshold algorithm to even \SI{66.5}{\percent} when using a fixed parameter. By using our \ac{fpr}-based approach, it can be increased to \SI{81.5}{\percent} and \SI{82.6}{\percent} respectively. However, there is still a large gap to the optimal parameter's accuracy. 

By combining both approaches this gap can be highly reduced and an accuracy of \SI{86.9}{\percent} is achieved, nearly approaching the optimal accuracies of \SI{89.4}{\percent} and \SI{88.7}{\percent}. Further, this model is the only one that is able to satisfy both application requirements without relying on an optimally adjusted parameter. Thus, only the combination of \ac{nn}-based and classical algorithms is able to satisfy our constraints in a practical application where the optimal pump-specific parameter is not known.  
\section{Conclusion \& Future Work}
\label{sec:conclusion}

In this work, we analyzed different algorithms for anomaly detection of industrial pumps.  In this context, we propose an \ac{ecnn} that receives additional inputs based on the statistics of the data and includes a pump-specific parameter to provide the required adaptability. For the design of the \ac{nn} we focus on low complexity to allow for the implementation on battery-powered edge devices. Further, we show how the pump-specific parameter can be determined and combine the \ac{ecnn} with a classical threshold algorithm to further increase the accuracy. As a result, this combined approach is the only one that satisfies both application requirements without relying on an optimal pump parameter. 

For future work, implementing the proposed algorithm on various edge devices, such as embedded \acp{gpu}, or \acp{fpga}, could provide valuable insights into power and energy consumption on real hardware. Additionally, analyzing different algorithms for determining the pump-specific parameter may lead to further optimization and improvement of the proposed method.

\bibliographystyle{IEEEtran}
\bibliography{IEEEabrv, bib_bibtex}

\end{document}